\documentclass[final]{rjparticle}
\usepackage{graphicx}

\DeclareSymbolFont{extraup}{U}{zavm}{m}{n}
\DeclareMathSymbol{\varheart}{\mathalpha}{extraup}{86}
\DeclareMathSymbol{\vardiamond}{\mathalpha}{extraup}{87}
\newcommand{\bdiamond}{{\scriptstyle \vardiamond}}
\newcommand{\cinf}{{{\cal \cC}^\infty(M,\R)}}

\newcommand{\bcE}{\boldsymbol{\check{E}}}

\newcommand{\vectornorm}[1]{||#1||}

\def\ad{\mathrm{ad}}
\newcommand{\D}{{\check{D}^\ad}}

\newcommand{\cone}{\mathrm{cone}}

\newcommand{\bp}{\begin{Proposition}}
\newcommand{\ep}{\end{Proposition}}
\newcommand{\bl}{\begin{Lemma}}
\newcommand{\el}{\end{Lemma}}
\newcommand{\bt}{\begin{Theorem}}
\newcommand{\et}{\end{Theorem}}
\newcommand{\bd}{\begin{Definition}}
\newcommand{\ed}{\end{Definition}}
\newcommand{\End}{\mathrm{End}}

\newcommand{\eqdef}{\stackrel{{\rm def.}}{=}}

\DeclareFontFamily{U}{rsf}{}
\DeclareFontShape{U}{rsf}{m}{n}{<5> <6> rsfs5 <7> <8> <9> rsfs7 <10-> rsfs10}{}
\DeclareMathAlphabet\Scr{U}{rsf}{m}{n}

\newcommand{\KA}{K\"{a}hler-Atiyah~}

\def\R{{\Bbb R}}

\def\dd{\mathrm{d}}

\def\vol{\mathrm{vol}}
\def\AdS{\mathrm{AdS}}

\newcommand{\be}{\begin{equation*}}
\newcommand{\ee}{\end{equation*}}
\newcommand{\ben}{\begin{equation}}
\newcommand{\een}{\end{equation}}
\newcommand{\beqa}{\begin{eqnarray*}}
\newcommand{\eeqa}{\end{eqnarray*}}
\newcommand{\beqan}{\begin{eqnarray}}
\newcommand{\eeqan}{\end{eqnarray}}

\newcommand{\nn}{\nonumber}

\def\cC{{\mathcal C}}

\def\cB{\Scr B}

\def\Cl{\mathrm{Cl}}

\def\cK{\mathrm{\cal K}}

\def\cC{\mathcal{C}}

\def\G_2{\mathrm{G_2}}

\title{Revisiting eight-‐manifold flux compactifications of M‐-theory using geometric algebra techniques} 
\author[1]{Elena-Mirela Babalic}
\author[1]{Calin Iuliu Lazaroiu}
\affil[1]{Department of Theoretical Physics,\\
``Horia Hulubei'' National Institute for Physics and Nuclear Engineering,\\
Reactorului 30, RO-077125, POB-MG6, Magurele-Bucharest, Romania\\
{\em Email}: mbabalic@theory.nipne.ro, lcalin@theory.nipne.ro}

\begin{document}
\nolinenumbers
\maketitle
\begin{abstract}
Motivated by open problems in F-theory, we reconsider warped
compactifications of M-theory on 8-manifolds to $\AdS_3$ spaces in the
presence of a non-trivial field strength of the M-theory 3-form, studying
the most general conditions under which such backgrounds preserve ${\cal N}=2$
supersymmetry in three dimensions. In contrast with previous studies, 
we allow for the most general pair of Majorana generalized Killing pinors 
on the internal 8-manifold, without imposing any chirality conditions on those pinors. 
We also show how such pinors can be  lifted to the 9‐-dimensional metric 
cone over the compactification 8-manifold.
We describe the translation of the generalized Killing pinor equations for such 
backgrounds to a system of differential and algebraic constraints on certain form-valued 
pinor bilinears and develop techniques through which such 
equations can be analyzed efficiently.
\end{abstract} 

\section{Introduction}
Consider eleven-dimensional supergravity on a background ${\tilde M}$ endowed
with a spinnable Lorentzian metric $\tilde g$ of `mostly plus' signature. The
fields of the theory are the three-form potential $\tilde C$ with four-form
field strength $\tilde G$, the gravitino $\tilde\Psi_M$ and the metric.  As in
\cite{MartelliSparks,Tsimpis}, we consider compactifications down to an
$\AdS_3$ space of cosmological constant $\Lambda=-8\kappa^2$, where $\kappa$
is a positive real parameter --- this includes the Minkowski case as the limit
$\kappa\rightarrow 0$.  Thus ${\tilde M}=N\times M$, where $N$ is an oriented
3-manifold diffeomorphic with $\R^3$ and carrying the $\AdS_3$ metric while
$M$ is an oriented Riemannian eight-manifold with metric denoted by $g$. The
metric ${\tilde g}$ is a warped product with the warp factor $\Delta$.  For
the field strength ${\tilde G}$, we use the ansatz:
\be
{\tilde G} = e^{3\Delta} G~~~{\rm with}~~~ G = {\rm vol}_3\wedge f+F~~,
\ee
where $f=f_m e^m\in \Omega^1(M)$, $F=\frac{1}{4!}F_{mnpq} e^{mnpq}\in \Omega^4(M)$ and
$\vol_3$ is the volume form of $N$. 
 
\noindent For the eleven-dimensional supersymmetry generator ${\tilde \eta}$, we
use the ansatz:
\be
{\tilde \eta}=e^{\frac{\Delta}{2}}\eta~~~{\rm with}~~~ \eta=\psi\otimes \xi~~,
\ee
where $\xi$ is a real pinor of spin $1/2$ on the internal space $M$ and $\psi$
is a real pinor on the $\AdS_3$ space $N$. As in \cite{MartelliSparks,
Tsimpis} (and in contradistinction with \cite{Becker}) {\em we do not require
that $\xi$ has definite chirality} \footnote{As we shall see in a moment, this seemingly trivial
generalization has drastic consequences, leading to a problem which is
technically much harder than that solved in the celebrated work of
\cite{Becker}.}. Mathematically, $\xi$ is a section of the
pinor bundle of $M$, which is a real vector bundle of rank $16$ defined on
$M$, carrying a fiberwise representation of the Clifford algebra $\Cl(8,0)$.
The corresponding morphism $\gamma:(\wedge T^\ast M,\diamond)\rightarrow
(\End(S),\circ)$ of bundles of algebras is an isomorphism. As in \cite{calin},
we have set  $\gamma^m\eqdef \gamma(e^m)$ and $\gamma^{(9)}\eqdef \gamma^1\circ \ldots \circ
\gamma^8$.
Assuming that $\psi$ is a Killing pinor on the $\AdS_3$ space, the
supersymmetry condition $\delta_{\tilde \eta}\tilde\Psi_M=0$ amounts to the following {\em constrained generalized
Killing (CGK) pinor equations} \cite{ga1} for $\xi$:
\ben
\label{par_eq}
D_m\xi = 0~~,~~Q\xi = 0~~,
\een
where $D_m$ is a linear connection on $S$ and $Q\in \Gamma(M,\End(S))$ is
a globally-defined endomorphism of the vector bundle $S$, given explicitly by: 
\beqan
\label{Dappl}
D_m=\nabla^S_m+A_m~~,~~A_m= \frac{1}{4}f_p\gamma_{m}{}^{p}\gamma^{(9)}
+\frac{1}{24}F_{m p q r}\gamma^{ p q r}+\kappa \gamma_m\gamma^{(9)}~~,\\
\label{Qappl}
Q=\frac{1}{2}\gamma^m\partial_m\Delta -\frac{1}{288}F_{m p q r}\gamma^{m p q r}
-\frac{1}{6}f_p \gamma^p \gamma^{(9)}
-\kappa\gamma^{(9)} ~~.
\eeqan
The space of solutions to \eqref{par_eq} is a finite-dimensional $\R$-linear
subspace $\cK(D,Q)$ of the space $\Gamma(M,S)$ of smooth globally-defined
sections of $S$. The problem of interest is to find those metrics and fluxes
on $M$ for which some fixed amount of supersymmetry is preserved in three
dimensions, i.e.  for which the space $\cK(D,Q)$ has some given non-vanishing
dimension, which we denote by $s$.  The case $s=1$ (which corresponds to ${\cal N}=1$
supersymmetry in three dimensions) was studied in \cite{MartelliSparks,
Tsimpis} and reconsidered in \cite{ga1} by using geometric algebra techniques.
The case $s=2$ (which leads to ${\cal N}=2$ supersymmetry in three dimensions) was
studied in \cite{Becker}, but considering only Majorana-{\em Weyl} solutions 
of \eqref{par_eq}, i.e. solutions $\xi$ which also satisfy the supplementary
constraint $\gamma^{(9)}\xi=\pm \xi$.  Here, we consider the case when no such
chirality constraint is imposed on the solutions of \eqref{par_eq}.

\section{The geometric algebra approach}
Since the procedure used is explained in detail in \cite{ga1, ga2} (being also summarized in \cite{calin})
in what follows we give only a brief overview.
\paragraph{The geometric product.}
Following an idea originally due to Chevalley and Riesz \cite{Chevalley,
Riesz}, we identify $\Cl(T^{\ast} M)$ with the exterior bundle $\wedge
T^{\ast} M$, thus realizing the Clifford product as the {\em geometric
product}, which is the fiberwise associative, unital and bilinear binary
composition $\diamond: \wedge T^{\ast} M \times_{M}\wedge T^{\ast} M
\rightarrow \wedge T^{\ast} M$ given on sections by the expansion:
\beqan
\label{starprod}
\omega\diamond \eta=\sum_{k=0}^{\left[\frac{d}{2}\right]}
\frac{(-1)^k}{(2k)!}\omega\wedge_{2k}\eta + \sum_{k=0}^{\left[\frac{d-1}{2}\right]} \frac{(-1)^{k+1}}{(2k+1)!}
\pi(\omega)\wedge_{2k+1}\eta~~,
\eeqan
where $\pi$ is the {\em grading automorphism} defined through:
\ben
\label{piDef}
\pi(\omega)\eqdef \sum_{k=0}^{d} (-1)^k\omega^{(k)} ~~,
~~\forall \omega=\sum_{k=0}^d\omega^{(k)}\in \Omega (M)~~,~~{\rm
  where}~~\omega^{(k)}\in \Omega^k (M)~~.
\een
The Clifford bundle is thus identified with the bundle of algebras $(\wedge
T^{\ast} M, \diamond)$, which is known as the {\em \KA bundle} of $(M,g)$.
The binary $\cinf$-bilinear operations $\wedge_k$ which appear in the
expansion above are the (action on sections of the) {\em contracted wedge
products}, defined iteratively through:
\be
\omega\wedge_0\eta \eqdef \omega\wedge \eta~~,~~\omega\wedge_{k+1}\eta
\eqdef g^{ab}(e_a \lrcorner \omega)\wedge_{k}(e_b\lrcorner \eta)=g_{ab}
(\iota_{e^a}\omega) \wedge_k (\iota_{e^b}\eta) ~~,
\ee
where $\iota$ denotes the so-called {\em interior product} (see \cite{ga1}).  We will mostly use, instead, the so-called {\em generalized products}
$\bigtriangleup_k$, which are defined by rescaling the contracted wedge products:
\ben
\label{GenProd}
\bigtriangleup_k\eqdef \frac{1}{k!}\wedge_k~~.
\een
 The \KA bundle also admits an involutive anti-automorphism
$\tau$ (known as {\em the main anti-automorphism} or as {\em
reversion}), which is given by:
\ben
\label{taudef}
\tau(\omega)\eqdef (-1)^{\frac{k(k-1)}{2}}\omega~~,~~\forall \omega\in
\Omega^k (M)~~.
\een

\section{Lifting the CGK equations to the metric cone over the
  compactification space}

The CGK pinor equations can be lifted to the metric cone $({\hat M}, g_\cone)$
over $M$ as explained in \cite{ga2} and outlined in \cite{calin},
to which we refer the reader for the notations used below.  As in loc. cit.,
we work with an admissible \cite{AC0,AC1} bilinear pairing $\cB$ on the pin
bundle $S$ of $M$ which is symmetric and with respect to which all $\gamma_m$ are self-adjoint.  We let ${\hat \cB}$
denote the pull-back of $\cB$ to the pin bundle ${\hat S}$ of the cone. We work with that pinor representation $\gamma_\cone$ on the cone which has
signature $+1$. 

\paragraph{The basic form-valued bilinears on the cone.} 
Recall that $s$ denotes the dimension of the space of solutions to the CGK pinor
equations. Choosing a basis $({\hat \xi}_i)_{i=1\ldots s}$ of such solutions
on the cone, we set (see \cite{ga1,ga2}):
\be
\check{E}^\cone_{ij}\eqdef  \check{E}^\cone_{{\hat \xi}_i,{\hat \xi}_j}=
\frac{1}{2^{\left[\frac{d+1}{2}\right]}}\bcE^\cone_{ij}\in \Omega^{+,\cone} ({\hat M})~~,
\ee
and:
\beqa
&&\bcE^\cone_{ij}=\sum_{k=0}^d\bcE^{(k),\cone}_{ij}=_U\sum_{k=0}^d \frac{1}{k!}
\bcE^{(k),\cone}_{a_1\ldots a_k}({\hat \xi}_i,{\hat \xi}_j) e^{a_1\ldots a_k}_{\cone,+}~~,\\
&& \bcE^{(k),\cone}_{a_1\ldots a_k}({\hat \xi}_i,{\hat \xi}_j)\eqdef {\hat
  \cB}(\gamma_{a_k\ldots a_1}{\hat \xi}_i, {\hat \xi}_j)={\hat \cB}({\hat \xi}_i, \gamma_{a_1\ldots a_k}{\hat \xi}_j)~~.
\eeqa
The symbol $\Omega^{+,\cone}({\hat  M})$ denotes the space of twisted self-dual forms on the cone , which is a
subalgebra of the \KA algebra of $({\hat M}, g_\cone)$ (see \cite{ga1, ga2} and \cite{calin}).  

\paragraph{CGK equations on the cone.}
As explained in \cite{ga2}, it is computationally convenient to replace the algebra $(\Omega^{+,\cone}({\hat M}), \diamond^\cone)$ of
twisted selfdual forms of the cone (which is the effective domain of
definition of $\gamma_\cone$) with a certain isomorphic
model $(\Omega^<({\hat M}), \bdiamond_+^\cone)$ whose precise definition is
given in loc. cit. When $s=2$ (${\cal N}=2$ supersymmetry in three dimensions), the CGK pinor equations on $\hat{M}$ admit {\em
two} linearly independent solutions $\xi_1$ and $\xi_2$.  We have
$\Omega^<({\hat M})=\oplus_{k=0}^{4}\Omega^k({\hat M})$, so we are interested
in pinor bilinears $\bcE^{(k)}_{{\hat \xi}_1,{\hat \xi}_2}$ with $k=0\ldots 4$
for two independent solutions ${\hat \xi}_1, {\hat \xi}_2\in \Gamma({\hat M},
{\hat S})$ of the CGK pinor
equations lifted to the cone (which are equivalent with the original CGK pinor equations on $M$):
\ben
\label{coneq}
{\hat D}_a{\hat \xi}={\hat Q}{\hat \xi}=0~~,
\een
where the definition of ${\hat D}_a=\nabla^{{\hat S},\cone}_a+A_a^\cone$,
where $a=1\ldots 9$, and ${\hat Q}$ can be found in \cite{ga2} and
\cite{calin}. 

To extract the translation of these equations into constraints on
differential forms, we implemented certain procedures within the
package {\tt Ricci} \cite{Ricci} for tensor computations in
{\tt Mathematica}\textsuperscript{\textregistered}. The dequantizations:
\beqa
\check{A}_a^\cone=\gamma_\cone^{-1}(A_a^\cone)\in \Omega^<({\hat M})~~,\\
\check{Q}^\cone =\gamma_\cone^{-1}({\hat Q})\in \Omega^<({\hat M})~~,
\eeqa
of $A^\cone$ and ${\hat Q}$ are given by $\check{A}_9^\cone=0$ and:
\begin{subequations}
\begin{align}
\check{A}_m^\cone =\frac{1}{4} \iota^\cone_{e_m^\cone}  F_\cone +
\frac{1}{4}(e_m^\cone)\wedge f_\cone \wedge \theta~~\in \Omega^<({\hat M})~~,
~~\forall m=1\ldots 8~~,\nn\\
\check{Q}^\cone=\frac{1}{2} r \dd
\Delta-\frac{1}{6}f_\cone \wedge\theta-\frac{1}{12}F_\cone -\kappa \theta ~~\in \Omega^<({\hat
M})~~,~~~~~~~~~~~~~~~~~~~~\nn
\end{align}
\end{subequations}
while the ${\hat \cB}$-transpose of ${\hat Q}$ dequantizes to the cone reversion of $\check{Q}^\cone$:
\be
{\hat \tau}(\check{Q}^\cone)=\frac{1}{2} r \dd \Delta+\frac{1}{6}f_\cone
\wedge\theta-\frac{1}{12}F_\cone -\kappa \theta ~~. \nn
\ee
The forms $f_\cone$ and $F_\cone$ above are the {\em cone lifts} (see
\cite{ga2}) of $f$ and $F$ respectively, while $\Delta$ stands for the
pullback $\Pi^\ast(\Delta)=\Delta\circ \Pi$ of the warp factor through the
natural projection $\Pi:{\hat M}\rightarrow M$, even though the notation does
not show this explicitly. The one-form $\theta$ is defined through:
\be
\theta\eqdef \dd r\in \Omega^1({\hat M})~~,
\ee 
where $r$ is the radial coordinate along the metric cone ${\hat M}$. 

A basis for the space spanned by the forms $\frac{1}{k!}{\hat \cB}({\hat \xi}_1, {\hat \gamma}_{a_1\ldots a_k}{\hat
\xi}_2)e^{a_1...a_k}_\cone\in \Omega^<({\hat M})$ (of rank $k\leq 4$) which can be constructed
on the cone from ${\hat \xi}_1$ and ${\hat \xi}_2$ is
given by (where we have raised all indices using the cone metric to avoid
notational clutter) :
\begin{subequations}
\begin{align}
&&V_1^a={\hat \cB}({\hat \xi}_1, {\hat \gamma}^a {\hat \xi}_1) ~~,~~ V_2^a ={\hat \cB}({\hat
\xi}_2, {\hat \gamma}^a {\hat \xi}_2) ~~,~~~ V_3^a = {\hat \cB}({\hat \xi}_1, {\hat \gamma}^a {\hat
\xi}_2)~,\nn\\
&&K^{ab} = {\hat \cB}({\hat \xi}_1, {\hat \gamma}^{ab} {\hat \xi}_2)~~,~~\Psi^{abc} =
{\hat \cB}({\hat \xi}_1, {\hat \gamma}^{abc} {\hat \xi}_2)~, \nn\\
&&\Phi_1^{abce}={\hat \cB}({\hat \xi}_1,{\hat \gamma}^{abce} {\hat \xi}_1) ~~,
~~ \Phi_2^{abce}={\hat \cB}({\hat \xi}_2,{\hat \gamma}^{abce} {\hat \xi}_2)
~~,~~\Phi_3^{abce} = {\hat \cB}({\hat \xi}_1,{\hat \gamma}^{abce} {\hat \xi}_2)~.\nn
\end{align}
To arrive at these bilinears we used the identity:
\be
\cB({\hat \xi}_i,{\hat \gamma}^{a_1\ldots a_k}{\hat \xi}_j)=(-1)^{\frac{k(k-1)}{2}}\cB({\hat \xi}_j,
{\hat \gamma} ^{a_1\ldots a_k}{\hat \xi}_i)~~,
\ee
which follows from the fact that $\gamma_a^t=\gamma_a$ and implies that certain of the forms
$\bcE^{(k),\cone}_{{\hat \xi}_i,{\hat \xi}_j}$ vanish identically.

\end{subequations}
Here and below, we have taken ${\hat \xi}_1$ and ${\hat \xi}_2$ to form a
${\hat \cB}$-orthonormal basis of the $\R$-vector space $\cK({\hat D},{\hat
  Q})$ of solutions to the CGK equations on the cone:
\be
{\hat \cB}({\hat \xi}_i,{\hat \xi}_j)=\delta_{ij}~~,~~\forall i,j=1,2~~,
\ee
and we noticed that the pairing ${\hat \cB}=\Pi^\ast(\cB)$ on ${\hat
  S}=\Pi^\ast(S)$ has the same symmetry and type
properties as the admissible pairing $\cB$ on $S$ --- namely, both $\cB$ and ${\hat \cB}$ are symmetric and
nondegenerate (and they can be taken to be positive-definite) and make the
eight- (respectively nine-) dimensional gamma `matrices' $\gamma^m$ and ${\hat
  \gamma}^a$ into self-adjoint operators.  From now on --- in order to avoid
notational clutter --- we shall suppress the superscripts and subscripts
``$\cone$''. In particular, we shall denote the cone lifts $F_\cone$ and
$f_\cone$ simply by $F$ and $f$. With these notations and conventions, the
truncated model $(\check{\cK}^{<,\cone}({\hat D},{\hat Q}),\bdiamond_+^\cone)$
of the flat Fierz algebra on the cone admits the basis:
\begin{subequations}
\begin{align}
&& \check{E}^<_{12} = \frac{1}{32}(V_3+K+\Psi+\Phi_3)~~,~~ \check{E}^<_{21} = \frac{1}{32}(V_3-K-\Psi+\Phi_3)~~, \nn\\
&&\check{E}^<_{11} = \frac{1}{32}(1+V_1+\Phi_1)~~,~~\check{E}^<_{22} =
\frac{1}{32}(1+V_2+\Phi_2)~~\nn
\end{align}
\end{subequations}
and can be generated by two elements (see Subsection 5.10 of \cite{ga1}), which we choose to be:
\be
\check{E}^<_{12}=\frac{1}{32}(V_3+K+\Psi+\Phi_3)~~,~~
\check{E}^<_{21}= {\hat \tau}(\check{E}^<_{12})=\frac{1}{32}(V_3-K-\Psi+\Phi_3)~. \nn
\ee
The overall coefficient $\frac{1}{32}$ comes from the prefactor
$\frac{1}{2^{\left[\frac{d+1}{2}\right]}}$ when $d=9$. 
As explained in \cite{ga1}, the Fierz relations for
the inhomogeneous forms $\check{E}_{ij}$ (which in this case are twisted selfdual $\check{E}_{ij}=\check{E}^<_{ij}+\tilde\ast\check{E}^<_{ij}$, with $\check{E}^<_{ij}\in\Omega^<(\hat{M})$)
 are given by:
\ben
\label{fierz}
\check{E}_{ij}\diamond \check{E}_{kl}= \delta_{jk}
\check{E}_{il}~~,~~\forall i,j,k,l=1,2~~.
\een
Since the associative and unital multiplication $\bdiamond_+$ is defined 
through:
\be
\omega\bdiamond_+\eta=2P_<(P_+(\omega)\diamond P_+(\eta)) ~\in\Omega^<(M)~~,
\ee
we find the following relations when taking $\omega=\check{E}^<_{ij}$ and $\eta=\check{E}^<_{kl}$ (with indices $i,j,k,l$ fixed):
\begin{subequations}
\begin{align}
\check{E}^<_{ij}\bdiamond_+\check{E}^<_{kl} =2P_<(P_+(\check{E}^<_{ij})\diamond P_+(\check{E}^<_{kl}))
 = 2P_<(\frac{1}{2}(\check{E}^<_{ij}+\tilde\ast\check{E}^<_{ij})\diamond (\frac{1}{2}(\check{E}^<_{kl}+\tilde\ast\check{E}^<_{kl}))\nn\\
 =\frac{1}{2}P_<(\check{E}_{ij}\diamond\check{E}_{kl})=\frac{1}{2} P_<(\delta_{jk}\check{E}_{il})=\frac{1}{2} \delta_{jk}\check{E}^<_{il}~~,~~~~~~~~~~~~~~~~~\nn
\end{align}
\end{subequations}
where we made use of \eqref{fierz}.  We thus obtain the Fierz
relations for the truncated model in our case:
\ben
\label{FierzRel}
\check{E}^<_{ij}\bdiamond_+\check{E}^<_{kl}=\frac{1}{2} \delta_{jk}\check{E}^<_{il}~~,~~\forall i,j,k,l=1,2~~.~~
\een
In order to avoid notational clutter, we shall henceforth use $\bdiamond$ 
instead of $\bdiamond_+$.

\

\noindent On the other hand, the algebraic constraints in \eqref{coneq} 
amount \cite{ga1} to the following two relations for $\check{E}^<_{12}$:
\ben
\label{AConstr}
 \check{Q}\bdiamond \check{E}^<_{12}\mp \check{E}^<_{12}\bdiamond {\hat \tau}(\check{Q})=0 ~~,
\een
while the differential constraints in \eqref{coneq} give 
$\D_a \check{E}^<_{12}=0\Longleftrightarrow \D_a
\check{E}^<_{21}=0$, which in turn imply:
\ben
\label{DConstr}
 \dd \check{E}^<_{12}= e^a\wedge\nabla_a \check{E}^<_{12}=-e^a\wedge[\check{A}_a,\check{E}^<_{12}]_{-,\bdiamond}~~.
\een
As explained in Subsection 5.10 of \cite{ga1}, it is enough to consider 
relations \eqref{AConstr} and \eqref{DConstr} for the generators
$\check{E}^<_{12}$ and $\check{E}^<_{21}={\hat \tau} (\check{E}^<_{12})$, since
the corresponding constraints for
$\check{E}^<_{11}=2\check{E}^<_{12}\bdiamond \check{E}^<_{21}$ and
$\check{E}^<_{22}=2\check{E}^<_{21}\bdiamond \check{E}^<_{12}$ follow from
those upon reversion.

\paragraph{Algebraic constraints.}
Using the procedures which we have implemented and the package {\tt Ricci}
\cite{Ricci} for tensor computations in {\tt
Mathematica}\textsuperscript{\textregistered}, we find that the first equation
(that with the minus sign) in \eqref{AConstr} amounts to the following system
when separated on ranks:

\begin{subequations}
\begin{align}
 \iota_{f\wedge\theta} K = 0 ~~,\nn\\
 r\iota_{\dd \Delta} K+\frac{1}{3}\iota_{f\wedge\theta}\Psi - \frac{1}{6}\iota_{\Psi} F -2 \kappa~\iota_\theta K  = 0 ~~,\nn\\
 \frac{1}{3} \iota_{f\wedge\theta}\Phi_3-\frac{1}{6}F\bigtriangleup_3 \Phi_3 +r(\dd \Delta)\wedge V_3+2\kappa~ V_3\wedge\theta = 0 ~~,\nn\\
 r\iota_{\dd \Delta} \Phi_3 -\frac{1}{3}V_3\wedge f\wedge\theta+\frac{1}{6}\iota_{V_3} F
 -\frac{1}{6}{\ast}(F\bigtriangleup_1 \Phi_3) +\frac{1}{3}{\ast}(f\wedge\theta\wedge \Phi_3 )-2\kappa~\iota_\theta\Phi_3 = 0 ~~,\nn\\
 r(\dd \Delta)\wedge\Psi-\frac{1}{3}f\wedge\theta\wedge K-\frac{1}{6} K\bigtriangleup_1 F
 -\frac{1}{3}{\ast}(f\wedge\theta\wedge\Psi) +\frac{1}{6}{\ast}(\Psi\bigtriangleup_1 F)+2\kappa~\Psi\wedge\theta = 0~~, \nn
\end{align}
\end{subequations}
while the second equation (that with the plus sign) in \eqref{AConstr} amounts to:
\begin{subequations}
\begin{align}
 -\frac{1}{6} \iota_F \Phi_3 +r\iota_{\dd \Delta} V_3 -2\kappa~\iota_\theta V_3 = 0 ~~,\nn\\
 \frac{1}{3} \iota_{V_3}(f\wedge\theta)-\frac{1}{6}{\ast}( F\wedge  \Phi_3 ) = 0 ~~,\nn\\
 r\iota_{\dd \Delta}\Psi +\frac{1}{3} (f\wedge\theta)\bigtriangleup_1 K+\frac{1}{6}  \iota_{K} F
 +\frac{1}{6}{\ast} (F\wedge\Psi)-2\kappa~ \iota_\theta \Psi = 0 ~~,\nn\\
 \frac{1}{3} (f\wedge\theta)\bigtriangleup_1\Psi +\frac{1}{6}\Psi\bigtriangleup_2 F +\frac{1}{6}{\ast}( K\wedge F)
 +r(\dd \Delta)\wedge K - 2\kappa~ K\wedge\theta = 0 ~~,\nn\\
 \frac{1}{3} (f\wedge\theta)\bigtriangleup_1 \Phi_3 + \frac{1}{6} F\bigtriangleup_2  \Phi_3
 -\frac{1}{6}{\ast}(F\wedge V_3) +{\ast}(r(\dd \Delta)\wedge \Phi_3 )-2\kappa\ast(\Phi_3\wedge\theta) = 0~~. \nn
\end{align}
\end{subequations}
\paragraph{Differential constraints.}
Using the same {\tt Mathematica}\textsuperscript{\textregistered} package, we
find that the differential constraints \eqref{DConstr}, when separated on ranks, amount to:
\begin{subequations}
\begin{align}
 \dd V_3 = \frac{1}{2} \Phi_3 \bigtriangleup_3 F + \iota_{f\wedge\theta} \Phi_3 ~~,  \nn\\
 \dd K = (f\wedge\theta)\bigtriangleup_1 \Psi + \Psi\bigtriangleup_2 F ~~,  \nn\\
 \dd \Psi = \frac{3}{2}F\bigtriangleup_1 K -\frac{1}{2}F\bigtriangleup_3{\ast}\Psi
 +2{\ast}(f\wedge\theta\wedge\Psi) -f\wedge\theta\wedge K ~~, \nn\\
 \dd \Phi_3  =-2 F\wedge V_3 +\frac{1}{2}e^m\wedge{\ast}((\iota_{e^m}F)\bigtriangleup_1 \Phi_3 )
  -\frac{1}{2}e^m\wedge{\ast}(((e_m) \wedge f\wedge\theta)\bigtriangleup_1 \Phi_3 ) ~~.\nn
\end{align}
\end{subequations}
According to our notational conventions, $e^m$ in the equations above stands
for the cone lifts $e^m_\cone$ etc. Furthermore, $\ast \eqdef \ast_\cone$ is
the ordinary Hodge operator of $(M,g_\cone)$ and $\iota$ stands for
$\iota^\cone$. The generalized products $\bigtriangleup_p\eqdef
\bigtriangleup_p^\cone$ are constructed with the cone metric on ${\hat M}$.

\paragraph{Fierz relations.}
Let us consider the Fierz identities \eqref{FierzRel} for the basis elements
$\check{E}^<_{ij}$ ($i,j=1,2$) of the truncated model of the flat
Fierz algebra on the cone:
\be
\begin{array}{ccccccc}
({\rm F1}):&~\check{E}^<_{12}\bdiamond\check{E}^<_{12}=0~~~~~~&,&({\rm F2}):&~\check{E}^<_{12}\bdiamond\check{E}^<_{21}=\frac{1}{2}\check{E}^<_{11}~~,\\
({\rm F3}):&~\check{E}^<_{12}\bdiamond\check{E}^<_{22}=\frac{1}{2}\check{E}^<_{12} &,&({\rm F4}):&~\check{E}^<_{12}\bdiamond\check{E}^<_{11}=0~~~~~~~~,\\
({\rm F5}):&~\check{E}^<_{11}\bdiamond\check{E}^<_{11}=\frac{1}{2}\check{E}^<_{11} &,&({\rm F6}):&~\check{E}^<_{11}\bdiamond\check{E}^<_{12}=\frac{1}{2}\check{E}^<_{12}~~,\\
({\rm F7}):&~\check{E}^<_{11}\bdiamond\check{E}^<_{21}=0~~~~~~&,&({\rm F8}):&~\check{E}^<_{11}\bdiamond\check{E}^<_{22}=0~~~~~~~~,\\
({\rm F9}):&~\check{E}^<_{21}\bdiamond\check{E}^<_{12}=\frac{1}{2}\check{E}^<_{22} &,&({\rm F10}):&~\check{E}^<_{21}\bdiamond\check{E}^<_{11}=\frac{1}{2}\check{E}^<_{21}~~,\\
({\rm F11}):&~\check{E}^<_{21}\bdiamond\check{E}^<_{21}=0~~~~~~&,&({\rm F12}):&~\check{E}^<_{21}\bdiamond\check{E}^<_{22}=0~~~~~~~~,\\
({\rm F13}):&~\check{E}^<_{12}\bdiamond\check{E}^<_{11}=0~~~~~~&,&({\rm F14}):&~\check{E}^<_{22}\bdiamond\check{E}^<_{12}=0~~~~~~~~,\\
({\rm F15}):&~\check{E}^<_{22}\bdiamond\check{E}^<_{21}=\frac{1}{2}\check{E}^<_{21}&,&({\rm F16}):&~\check{E}^<_{22}\bdiamond\check{E}^<_{22}=\frac{1}{2}\check{E}^<_{22}~~.
\end{array}
\ee
We note that some of these conditions are equivalent through reversion
(namely (F1)$\Leftrightarrow$(F11), (F3)$\Leftrightarrow$(F15),
(F4)$\Leftrightarrow$(F7), (F6)$\Leftrightarrow$(F10),
(F8)$\Leftrightarrow$(F13) and (F12)$\Leftrightarrow$(F14)), while
relations (F2), (F5), (F9), (F16) are selfdual under reversion. Expanding for example equation (F1):
\be
(V_3 + K + \Psi + \Phi_3)\bdiamond (V_3 + K + \Psi + \Phi_3)=0
\ee
gives the following conditions when separating rank components:
\begin{subequations}
\begin{align}
-\vectornorm{K}^2 + || \Phi_3||^2 - || \Psi ||^2 + || V_3 ||^2 = 0
~~, \nn\\ -2\iota_K\Psi +{\ast}(\Phi_3\wedge\Phi_3) = 0 ~~, \nn\\
\iota_{V_3} \Psi - {\ast}(\Phi_3\wedge\Psi)-\iota_K \Phi_3 =0
~~, \nn\\ K\wedge V_3 -{\ast}(K\wedge\Phi_3) -
\Psi\bigtriangleup_2\Phi_3 = 0 ~~, \nn\\ \Psi\bigtriangleup_1\Psi
-\Phi_3\bigtriangleup_2\Phi_3 + 2{\ast}(K\wedge\Psi) +
2{\ast}(V_3\wedge\Phi_3) +K\wedge K = 0~~. \nn
\end{align}
\end{subequations}
Similarly, all independent Fierz relations given above can be expanded into
rank components and studied by elimination. Such a detailed analysis and various
implications are taken up in a forthcoming publication.

\section{Conclusions}
The geometric algebra approach developed in \cite{ga1,ga2} provides a
synthetic and computationally efficient method for translating generalized
Killing (s)pinor equations into conditions on differential forms constructed
as (s)pinor bilinears. This approach is highly amenable 
to implementation in various symbolic computation systems specialized 
in tensor algebra --- and we touched upon two such implementations which we have
carried out using {\tt Ricci} \cite{Ricci}. It affords a more unified and systematic description of flux 
compactifications and generally of supergravity and string compactifications.
We illustrated our techniques with the most general flux
compactifications of M-theory preserving ${\cal N}=2$ supersymmetry in
three dimensions, a class of compactifications which had not been studied 
in full generality before --- showing on the one hand how to obtain a complete description
of the differential and algebraic constraints on pinor bilinears and on the other hand how to write
all Fierz identities between the form bilinears.  A detailed analysis of
the resulting equations, geometry and physics is the subject of
ongoing work. The methods introduced in \cite{ga1,ga2} have much wider applicability, leading to 
promising new directions in the study of supergravity and string theory
backgrounds and actions.

\begin{acknowledgement}
This work was supported by the CNCS projects PN-II-RU-TE (contract number
77/2010) and PN-II-ID-PCE (contract numbers 50/2011 and 121/2011). The authors
thank the organizers of the 8-th QFTHS Conference for hospitality and interest
in their work.  C.I.L thanks the Center for Geometry and Physics, Institute
for Basic Science and Pohang University of Science and Technology (POSTECH),
Pohang, Korea for providing excellent conditions at various stages during the
preparation of this work, through the research visitor program affiliated with
Grant No. CA1205-1. The Center for Geometry and Physics is supported by the
Government of Korea through the Research Center Program of IBS (Institute for
Basic Science).  He also thanks Perimeter Institute for hospitality for
providing an excellent and stimulating research environment during the last
stages of this project. Research at Perimeter Institute is
supported by the Government of Canada through Industry Canada and by the
Province of Ontario through the Ministry of Economic Development and
Innovation.
\end{acknowledgement}

\end{document}